\documentclass[12pt]{amsart}
\usepackage[latin1]{inputenc}
\usepackage{amssymb}
\usepackage{a4wide}
\usepackage{graphicx}
\usepackage{verbatim}
\usepackage{amsmath}
\usepackage{version}
 \theoremstyle{plain}    
 \newtheorem{thm}{Theorem}[section]
 \numberwithin{equation}{section} %% Comment out for sequentially-numbered
 \numberwithin{figure}{section} %% Comment out for sequentially-numbered
 \theoremstyle{plain}
 \theoremstyle{plain}    
 \theoremstyle{plain}    
 \theoremstyle{plain}    
 \newtheorem{conj}[thm]{Conjecture} %%Delete [thm] to re-start numbering
\theoremstyle{definition}
\newtheorem{rem}[thm]{Remark} %%Delete [thm] to re-start numbering
\theoremstyle{definition}
\newtheorem{definition}{Definition}
\newcommand{\Lim}{\mathop{\longrightarrow}\limits}

\begin{document}
\newcommand{\nwc}{\newcommand}
\nwc{\nwt}{\newtheorem}
\nwt{coro}{Corollary}
\nwt{ex}{Example}

%font change

\nwc{\mf}{\mathbf} %Latex (as in \bf not tilted math letters)
\nwc{\blds}{\boldsymbol} %Latex 
\nwc{\ml}{\mathcal} %Latex

%greek letters

\nwc{\lam}{\lambda}
\nwc{\del}{\delta}
\nwc{\Del}{\Delta}
\nwc{\Lam}{\Lambda}
\nwc{\elll}{\ell}
%blackboard bold math

\nwc{\IA}{\mathbb{A}} %algebraic
\nwc{\IB}{\mathbb{B}} %ball
\nwc{\IC}{\mathbb{C}} %complex
\nwc{\ID}{\mathbb{D}} %Dedekind
\nwc{\IE}{\mathbb{E}} %Euklides
\nwc{\IF}{\mathbb{F}} %finite field
\nwc{\IG}{\mathbb{G}} %Gauss
\nwc{\IH}{\mathbb{H}} %Hilbert\N-subgroup
\nwc{\IN}{\mathbb{N}} %natural
\nwc{\IP}{\mathbb{P}} %prime
\nwc{\IQ}{\mathbb{Q}} %rational
\nwc{\IR}{\mathbb{R}} %real
\nwc{\IS}{\mathbb{S}} %sphere
\nwc{\IT}{\mathbb{T}} %torus
\nwc{\IZ}{\mathbb{Z}} %integers
\def\bbbone{{\mathchoice {1\mskip-4mu {\rm{l}}} {1\mskip-4mu {\rm{l}}}
{ 1\mskip-4.5mu {\rm{l}}} { 1\mskip-5mu {\rm{l}}}}}
\def\bbleft{{\mathchoice {[\mskip-3mu {[}} {[\mskip-3mu {[}}{[\mskip-4mu {[}}{[\mskip-5mu {[}}}}
\def\bbright{{\mathchoice {]\mskip-3mu {]}} {]\mskip-3mu {]}}{]\mskip-4mu {]}}{]\mskip-5mu {]}}}}
\nwc{\setK}{\bbleft 1,K \bbright}
\nwc{\setN}{\bbleft 1,\cN \bbright}
%Straight (vector) bold letters

%lowercase

\nwc{\va}{{\bf a}}
\nwc{\vb}{{\bf b}}
\nwc{\vc}{{\bf c}}
\nwc{\vd}{{\bf d}}
\nwc{\ve}{{\bf e}}
\nwc{\vf}{{\bf f}}
\nwc{\vg}{{\bf g}}
\nwc{\vh}{{\bf h}}
\nwc{\vi}{{\bf i}}
\nwc{\vI}{{\bf I}}
\nwc{\vj}{{\bf j}}
\nwc{\vk}{{\bf k}}
\nwc{\vl}{{\bf l}}
\nwc{\vm}{{\bf m}}
\nwc{\vM}{{\bf M}}
\nwc{\vn}{{\bf n}}
\nwc{\vo}{{\it o}}
\nwc{\vp}{{\bf p}}
\nwc{\vq}{{\bf q}}
\nwc{\vr}{{\bf r}}
\nwc{\vs}{{\bf s}}
\nwc{\vt}{{\bf t}}
\nwc{\vu}{{\bf u}}
\nwc{\vv}{{\bf v}}
\nwc{\vw}{{\bf w}}
\nwc{\vx}{{\bf x}}
\nwc{\vy}{{\bf y}}
\nwc{\vz}{{\bf z}}
\nwc{\bal}{\blds{\alpha}}
\nwc{\bep}{\blds{\epsilon}}
\nwc{\barbep}{\overline{\blds{\epsilon}}}
\nwc{\bnu}{\blds{\nu}}
\nwc{\bmu}{\blds{\mu}}
\nwc{\bet}{\blds{\eta}}

%bold letters
%\b* letters are tilted in math mode and scale in equations. 
%but cannot be used in plain text format.

%I. lowercase

\nwc{\bk}{\blds{k}}
\nwc{\bm}{\blds{m}}
\nwc{\bM}{\blds{M}}
\nwc{\bp}{\blds{p}}
\nwc{\bq}{\blds{q}}
\nwc{\bn}{\blds{n}}
\nwc{\bv}{\blds{v}}
\nwc{\bw}{\blds{w}}
\nwc{\bx}{\blds{x}}
\nwc{\bxi}{\blds{\xi}}
\nwc{\by}{\blds{y}}
\nwc{\bz}{\blds{z}}

%caligraphic

\nwc{\cA}{\ml{A}}
\nwc{\cB}{\ml{B}}
\nwc{\cC}{\ml{C}}
\nwc{\cD}{\ml{D}}
\nwc{\cE}{\ml{E}}
\nwc{\cF}{\ml{F}}
\nwc{\cG}{\ml{G}}
\nwc{\cH}{\ml{H}}
\nwc{\cI}{\ml{I}}
\nwc{\cJ}{\ml{J}}
\nwc{\cK}{\ml{K}}
\nwc{\cL}{\ml{L}}
\nwc{\cM}{\ml{M}}
\nwc{\cN}{\ml{N}}
\nwc{\cO}{\ml{O}}
\nwc{\cP}{\ml{P}}
\nwc{\cQ}{\ml{Q}}
\nwc{\cR}{\ml{R}}
\nwc{\cS}{\ml{S}}
\nwc{\cT}{\ml{T}}
\nwc{\cU}{\ml{U}}
\nwc{\cV}{\ml{V}}
\nwc{\cW}{\ml{W}}
\nwc{\cX}{\ml{X}}
\nwc{\cY}{\ml{Y}}
\nwc{\cZ}{\ml{Z}}

%% (wide)tilde letters

\nwc{\tA}{\widetilde{A}}
\nwc{\tB}{\widetilde{B}}
\nwc{\tE}{E^{\vareps}}
%\nwc{\tcO}{\widetilde{\mathcal{O}}}
\nwc{\tk}{\tilde k}
\nwc{\tN}{\tilde N}
\nwc{\tP}{\widetilde{P}}
\nwc{\tQ}{\widetilde{Q}}
\nwc{\tR}{\widetilde{R}}
\nwc{\tV}{\widetilde{V}}
\nwc{\tW}{\widetilde{W}}
\nwc{\ty}{\tilde y}
\nwc{\teta}{\tilde \eta}
\nwc{\tdelta}{\tilde \delta}
\nwc{\tlambda}{\tilde \lambda}
%\nwc{\tchi}{\tilde \chi}
\nwc{\ttheta}{\tilde \theta}
\nwc{\tvartheta}{\tilde \vartheta}
\nwc{\tPhi}{\widetilde \Phi}
\nwc{\tpsi}{\tilde \psi}
\nwc{\tmu}{\tilde \mu}

%miscellany
\nwc{\To}{\longrightarrow} %limits

\nwc{\ad}{\rm ad}
\nwc{\eps}{\epsilon}
\nwc{\ep}{\epsilon}
\nwc{\vareps}{\varepsilon}

\def\ep{\epsilon}
\def\tr{{\rm tr}}
\def\Tr{{\rm Tr}}
\def\i{{\rm i}}
\def\mi{{\rm i}}
\def\e{{\rm e}}
\def\sq2{\sqrt{2}}
\def\sqn{\sqrt{N}}
\def\vol{\mathrm{vol}}
\def\defi{\stackrel{\rm def}{=}}
\def\t2{{\mathbb T}^2}
%\def\tt2{{\mathbb T}^2}
%\nwc{\t1}{{\mathbb T}^1}
\def\s2{{\mathbb S}^2}
\def\hn{\mathcal{H}_{N}}
\def\shbar{\sqrt{\hbar}}
\def\A{\mathcal{A}}
\def\N{\mathbb{N}}
\def\T{\mathbb{T}}
\def\R{\mathbb{R}}
\def\RR{\mathbb{R}}
\def\Z{\mathbb{Z}}
\def\C{\mathbb{C}}
\def\O{\mathcal{O}}
\def\Sp{\mathcal{S}_+}
\def\Lap{\triangle}
\nwc{\lap}{\bigtriangleup}
\nwc{\rest}{\restriction}
\nwc{\Diff}{\operatorname{Diff}}
\nwc{\diam}{\operatorname{diam}}
\nwc{\Res}{\operatorname{Res}}
\nwc{\Spec}{\operatorname{Spec}}
\nwc{\Vol}{\operatorname{Vol}}
\nwc{\Op}{\operatorname{Op}}
\nwc{\supp}{\operatorname{supp}}
\nwc{\Span}{\operatorname{span}}

\nwc{\dia}{\varepsilon}
\nwc{\cut}{f}
\nwc{\qm}{u_\hbar}

\def\hto0{\xrightarrow{\hbar\to 0}}
\def\htoo{\stackrel{h\to 0}{\longrightarrow}}
\def\rto0{\xrightarrow{r\to 0}}
\def\rtoo{\stackrel{r\to 0}{\longrightarrow}}
\def\ntoinf{\xrightarrow{n\to +\infty}}

\providecommand{\abs}[1]{\lvert#1\rvert}
\providecommand{\norm}[1]{\lVert#1\rVert}
\providecommand{\set}[1]{\left\{#1\right\}}

\nwc{\la}{\langle}
\nwc{\ra}{\rangle}
\nwc{\lp}{\left(}
\nwc{\rp}{\right)}

%\nwc{\bal}{\begin{align}}
\nwc{\bequ}{\begin{equation}}
\nwc{\be}{\begin{equation}}
\nwc{\ben}{\begin{equation*}}
\nwc{\bea}{\begin{eqnarray}}
\nwc{\bean}{\begin{eqnarray*}}
\nwc{\bit}{\begin{itemize}}
\nwc{\bver}{\begin{verbatim}}

%\nwc{\eal}{\end{align}}
\nwc{\eequ}{\end{equation}}
\nwc{\ee}{\end{equation}}
\nwc{\een}{\end{equation*}}
\nwc{\eea}{\end{eqnarray}}
\nwc{\eean}{\end{eqnarray*}}
\nwc{\eit}{\end{itemize}}
\nwc{\ever}{\end{verbatim}}

\newcommand{\defeq}{\stackrel{\rm{def}}{=}}

\title[Chaotic vibrations]
{Chaotic vibrations and strong scars}

\author[N. Anantharaman]{Nalini Anantharaman}
\author[S. Nonnenmacher]{St\'ephane Nonnenmacher}
\address{CMLS, \'Ecole Polytechnique, 91128 Palaiseau, France}
\email{nalini@math.polytechnique.fr}
\address{Institut de Physique Th\'eorique, 
CEA/DSM/IPhT, Unit\'e de recherche associ\'ee au CNRS,
CEA/Saclay, 91191 Gif-sur-Yvette, France}
\email{snonnenmacher@cea.fr}

%\begin{abstract}

%\end{abstract}
%Dans sa célèbre formule ``Peut-on entendre la forme d'un tambour?'',  Mark Kac
%fait allusion au lien profond entre la géométrie du tambour (ou d'une plaque vibrante) et
%le spectre de ses modes stationnaires de vibration. 

\maketitle

\section{Introduction}

What relates an earthquake, a drum, a 2-dimensional mesoscopic cavity, a microwave oven and
an optical fibre? The equations governing wave propagation in these systems (seismic,
acoustic, electronic, microwave and optical) are linear. As a result, the solutions of the
wave equations can be decomposed as a sum over vibrating eigenmodes (or stationary modes).
The discrete (or ``quantum'') nature of this eigenmode decomposition is due to the compact
geometry of the above-mentioned ``cavities''. Mathematically, the latter are modelled by
compact, $d$-dimensional Riemannian manifolds $(X,g)$ with or without boundaries.

To simplify the presentation, we will restrict ourselves to scalar waves, described by a
real wavefunction $\psi(x,t)$. The eigenmodes $(\psi_n(x))_{n\geq 0}$ then satisfy
Helmholtz's equation
\begin{equation}\label{e:Helmholtz}
\Delta \psi_n + k_n^2\,\psi_n =0\,,
\end{equation}
where $\Delta:H^2\to L^2$ is the Laplace-Beltrami operator on $X$ and $k_n\geq 0$ is the
vibration frequency of the mode $\psi_n$. If the manifold has a boundary (as is the case
for an acoustic drum or for electromagnetic cavities), the wavefunction must generally
satisfy specific boundary conditions, dictated by the physics of the system: the simplest
ones are the Dirichlet ($\psi_{|\partial X}=0$) and Neumann ($\partial_{\nu}\psi_{|\partial
X}=0$) boundary conditions, where $\partial_\nu$ is the normal derivative at the boundary.

Our goal is to describe the eigenmodes, in particular the high-frequency eigenmodes
($k_n\gg 1$). Specifically, we would like to predict the localization properties of the
modes $\psi_n$, from our knowledge of the geometry of the manifold $(X,g)$.

Consider, for instance, the case of a bounded domain in the Euclidean plane, which we will
call a \textit{billiard}. For some very particular billiard shapes (e.g.\ a rectangle, a
circle or an ellipse), there exists a choice of coordinates allowing one to separate the
variables in Helmholtz's equation \eqref{e:Helmholtz}, thereby reducing it to a
one-dimensional eigenvalue problem (of the Sturm-Liouville type). In the high-frequency
limit, the latter can be solved to arbitrarily high precision through
WKB\footnote{Wentzel-Kramers-Brillouin.} methods, or sometimes even exactly (see Figure
\ref{fig:cercle}). The high-energy eigenmodes of such domains are hence very
well-\linebreak understood.

% \begin{figure*}
% \centering
% \includegraphics[angle=-90,width=0.29\textwidth]{anantharaman-orbit_kreis}$\ $
% \includegraphics[angle=-90,width=0.6\textwidth]{anantharaman-circle37ee-bis}
% 
% \caption{Left: one orbit of the circular billiard. Centre and right: two eigenmodes of that
% billiard, with their respective frequencies. }\label{fig:cercle} 
% 
% \end{figure*}

This separation of variables can be interpreted as a particular symmetry of the
\textit{classical dynamics} of the billiard (the motion of a point particle rolling
frictionless across the billiard and bouncing on its boundary). This dynamics is
\textit{Liouville-integrable}, which means that there is a conserved quantity in addition
to the kinetic energy. For instance, in a circular billiard the angular momentum of the
particle is conserved. The classical trajectories are then very ``regular'' (see Figure
\ref{fig:cercle}). The same regularity is observed in the eigenfunctions and can be
explained by the existence of a non-trivial differential operator commuting with the
Laplacian.
\begin{figure}
\begin{center}
\includegraphics[angle=-90,width=0.29\textwidth]{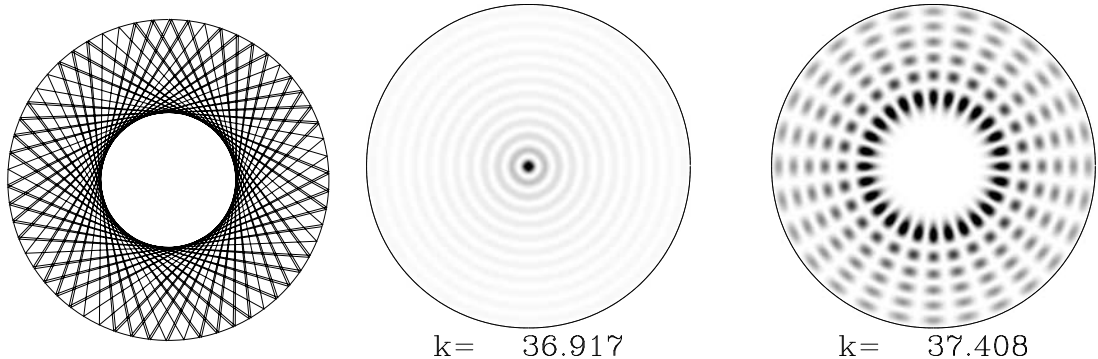}$\ $
\includegraphics[angle=-90,width=0.6\textwidth]{circle37ee-bis.ps}
\caption{\label{fig:cercle} Left: one orbit of the circular billiard. Center and right: two eigenmodes of that
billiard, with their respective frequencies. }
\end{center}
\end{figure}
As soon as an integrable billiard is slightly deformed, the symmetry is broken: the
geodesic flow is no longer integrable; it becomes \textit{chaotic} in some regions of phase
space. We do not have any approximate formula at hand to describe the eigenmodes. The
extreme situation consists of \textit{fully chaotic} billiards, like the ``stadium''
displayed in Figure \ref{fig:stade} (the word ``chaotic'' is a fuzzy notion; the results we
present below will always rely on precise mathematical assumptions).

We mention that the most recent numerical methods (the boundary operator and the ``scaling
method'') allow one to compute a few tens of thousands of eigenmodes for 2-dimen\-sional
billiards, at most a few thousands in 3 dimensions and much less if the metric is not
Euclidean. The difficulty stems from the fact that a mode of frequency $k_n\gg 1$
oscillates on a scale $\sim 1/k_n$ (the wavelength); one thus needs a finer and finer mesh
when increasing the frequency\footnote{The code used to compute the stadium eigenmodes
featured in this article was written and provided by Eduardo Vergini \cite{VS95}.}. On the
other hand, the analytical methods and results we present below are especially fitted to
describe these high-frequency modes.

\begin{figure}
\begin{center}
\includegraphics[angle=-90,width=.4\textwidth]{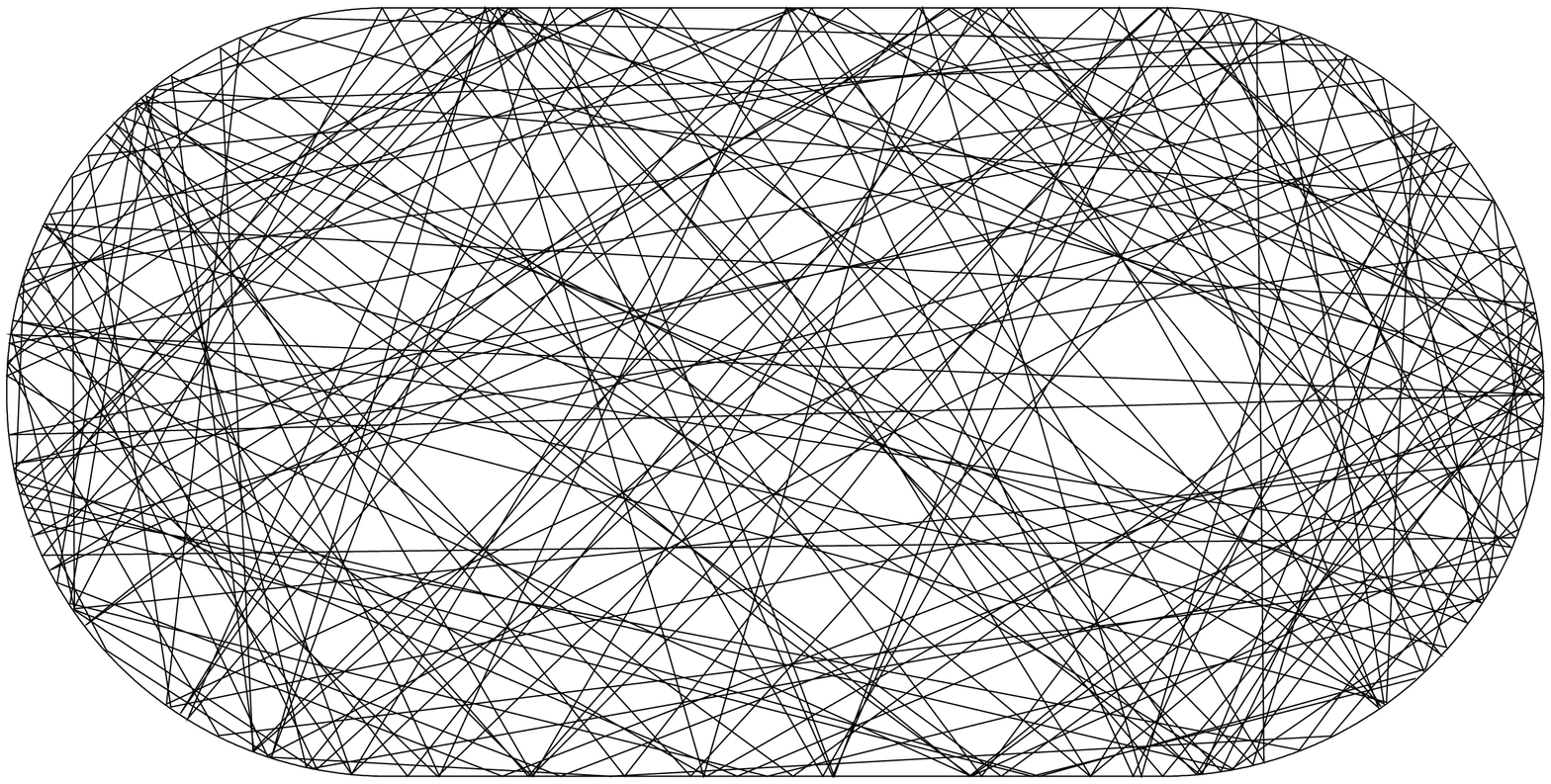}\hspace{.5cm}
\includegraphics[angle=-90,width=.4\textwidth]{stadium39ee-ergod.ps}\\
\includegraphics[angle=-90,width=.4\textwidth]{stadium39ee-scar.ps}\hspace{.5cm}
\includegraphics[angle=-90,width=.4\textwidth]{stadium39ee-bb.ps}
\caption{\label{fig:stade} Top left: one typical ``ergodic'' orbit of the ``stadium'': 
it equidistribues across the whole
billiard.  The three other plots feature eigenmodes of frequencies $k_n\approx 39$. Bottom left: a ``scar'' on
the (unstable) horizontal periodic orbit. Bottom right: a ``bouncing ball'' mode.}
\end{center}
\end{figure}

\subsection{Semiclassical methods}

In the general case of a Riemannian manifold, the classical dynamics (away from the
boundaries) consists of the Hamiltonian flow $g^t$ on the cotangent bundle\footnote{This
bundle is often called ``phase space''. It consists of the pairs $(x,\xi)$, where $x\in X$
and $\xi\in\IR^d$ is the coordinate of a covector based at $x$, representing the momentum
of the particle.} $T^*X$, generated by the free motion Hamiltonian
\begin{equation}\label{e:hamiltonien}
H(x,\xi)=\frac{|\xi|^2}2, \quad (x,\xi)\in T^* X.
\end{equation}
The flow on the energy layer $H^{-1}(1/2)=S^*X=\{(x,\xi)\,:\,|\xi|=1\}$ is simply the
\textit{geodesic flow} on the manifold (with reflections on the boundary in the case
$\partial X\neq 0$).

The high-frequency regime allows us to use the tools of semiclassical analysis. Indeed, the
Helmholtz equation \eqref{e:Helmholtz} can be interpreted as a stationary Schr\"odinger
equation: taking $\hbar_n=k_n^{-1}$ as an ``effective Planck's constant'', the eigenmode
$\psi_n$ satisfies
\begin{equation}\label{e:Helmholtz2}
-\frac{\hbar_{n}^2 \Delta}{2}\, \psi_n = \frac12\,\psi_n.
\end{equation}
The operator $-\frac{\hbar^2 \Delta}{2}$ on the left-hand side is the quantum Hamiltonian
governing the dynamics of a particle moving freely inside the cavity; it is the
\textit{quantization} of the classical Hamiltonian \eqref{e:hamiltonien}. The above
equation describes a quantum particle in a stationary state of energy $E=1/2$ (in this
formalism, the energy is fixed but Planck's ``constant'' is the running variable). The
high-frequency limit $k_n\to\infty$ exactly corresponds to the semiclassical regime
$\hbar=\hbar_{n}\to 0$. In the following, the eigenmode will be denoted by $\psi_n$ or
$\psi_\hbar$.

The \textit{correspondence principle} provides a connection between the
\textit{Schr\"odinger propagator}, namely the unitary flow
$U^t=e^{{it\hbar}\frac{\Delta}2}$ acting on $L^2(X)$ and the geodesic flow $g^t$ acting on
the phase space $T^*X$. The former ``converges'' towards the latter in the semiclassical
limit $\hbar\to 0$, in a sense made explicit below. The aim of semiclassical analysis is to
exploit this correspondence and use our understanding of the geodesic flow in order to
extract properties of the Schr\"odinger flow.

To analyse the eigenmodes we need to \textit{observe} them by using \textit{quantum
observables}. For us, an observable is a real function $A\in C^\infty(T^*X)$ that will be
used as a test function to measure the phase space localization of a wavefunction. One can
associate to this function a quantum observable $\Op_\hbar(A)$, which is a selfadjoint
operator on $L^2(X)$ obtained from $A$ through a certain ($\hbar$-dependent) quantization
procedure. For instance, on $X=\IR^d$ a possible procedure is the Weyl quantization
\begin{equation}\label{e:Weyl}
\Op_\hbar^W(A)f(x) =\frac{1}{(2\pi\hbar)^d} \int A\left(\frac{x+y}2, \xi\right) e^{\frac{i}\hbar \xi.(x-y)}f(y)dy\, d\xi.
\end{equation}
The simplest case consists of functions $A(x,\xi)=A(x)$ independent of the momentum;
$\Op_\hbar^W(A)$ is then the operator of multiplication by $A(x)$. If $A=A(\xi)$ is a
polynomial in the variable $\xi$ then $\Op_\hbar^W(A)$ is the differential operator
\smash{$A\left(\frac\hbar{i}\frac{\partial}{\partial x}\right)$}. The role of the parameter
$\hbar$ in the definition of $\Op_\hbar^W(A)$ is to adapt that operator to the study of
functions oscillating on a spatial scale $\sim\hbar$. On a general smooth manifold $X$, one
can define a quantization $\Op_\hbar(A)$ by using the formula \eqref{e:Weyl} in local
charts and then glue together the charts using a smooth partition of unity. 

%This procedure of quantization is not unique. But two distinct quantizations
%will enjoy the same properties in the semiclassical limit.

A mathematical version of the correspondence principle takes the form of an \emph{Egorov
theorem}. It states that quantization (approximately) commutes with evolution for
observables:
\begin{equation}\label{e:egorov}
\Vert e^{-it\hbar \frac{\Delta}2  }\Op_\hbar(A)e^{{it\hbar}\frac{\Delta}2}-\Op_\hbar(A\circ g^t)\Vert_{\cL(L^2)} =
\cO_{A,t}(\hbar)\,.
\quad\text{{(Egorov)}} 
\end{equation}

\subsection{Semiclassical measures}

\begin{sloppy}

In quantum mechanics, the function $|\psi(x)|^2$ describes the \linebreak probability
(density) of finding the particle at the position $x\in X$. A measuring device will only be
able to measure the probability integrated over a small region (a ``pixel'') $\int_B
|\psi(x)|^2\,dx$, which can be expressed as a diagonal matrix element $\la \psi, \bbbone_B
\psi\ra$. Here $\bbbone_B$ is the multiplication operator (on $L^2(X)$) by the
characteristic function on $B$.

\end{sloppy}

More generally, for a nontrivial observable $A(x,\xi)$ supported in a small \textit{phase
space} region, the matrix element  $\la \psi,\Op_\hbar(A)\psi\ra$ provides
information on the probability of the particle lying in this region. From the linearity of
the quantization scheme $A\mapsto \Op_\hbar(A)$, this matrix element defines a distribution
$\mu_\psi$ on $T^*X$:
\[
\mu_\psi(A)\defeq \la \psi,\Op_\hbar(A)\psi\ra,\qquad \forall A\in C_0^\infty(T^*X).
\]
This distribution (which depends on the state $\psi$ but also on the scale $\hbar$) is 
called the \textit{Wigner measure of the state $\psi$}.
%in spite of the fact that this measure
%is generally not positive. 
The projection of $\mu_\psi$ on $X$ is equal to the probability measure $|\psi(x)|^2\,dx$;
for this reason, $\mu_\psi$ is also called a \textit{microlocal lift} of that measure.
Still, $\mu_\psi$ contains more information: it takes the \textit{phase} of $\psi$ into
account and thereby also describes the local momentum of the particle (measured at the
scale $\hbar$).
%\footnote{The full information on the state $\psi$ (apart from a global phase factor)
%is encoded in the measure $\mu_\psi$.}. 

In order to study the localization properties of the eigenmodes $\psi_n$, we will consider
their Wigner measures $\mu_{\psi_n}=\mu_n$ (constructed with the adapted scales $\hbar_n$).
It is difficult to state anything rigorous about the Wigner measures of individual
eigenmodes so we will only aim to understand the limits of (subsequences of) the family
$(\mu_n)_{n\geq 0}$ in the weak topology on distributions. Such a limit $\mu$ is called a
\textbf{semiclassical measure} of the manifold $X$. Basic properties of the quantization
scheme imply that:
\begin{itemize}%[--]
\item $\mu$ is a probability measure supported on the energy shell $S^*X$.
% (the only important fact being that $\Op_\hbar(A)$ is a pseudodifferential operator with principal symbol $A$),
\item $\mu$ is \textit{invariant} through the geodesic flow:  $\mu=(g^{t})_*(\mu)$, $\forall t\in\IR$.
\item The collection of semiclassical measures $\mu$ does not depend on the choices of local symplectic coordinates involved in the definition of the quantization scheme $A\mapsto\Op_\hbar(A)$.
\end{itemize}
The second property is a direct consequence of the Egorov theorem \eqref{e:egorov}.

Starting from the family of quantum stationary modes $(\psi_{n})_n$, we have
constructed one or several probability measures $\mu$ on $S^*X$, invariant through the
classical flow. Each of them describes the asymptotical localization properties of the
modes in some subsequence $(\psi_{n_j})_{j\geq 1}$.

\vspace{-.1\baselineskip}

\subsection{Is any invariant measure a semiclassical measure?}

On a general Riemannian manifold $X$, the geodesic flow admits many different invariant
probability measures. The Liouville measure, defined as the disintegration on the energy
shell $S^*X$ of the symplectic volume $dx\,d\xi$, is a ``natural'' invariant measure on
$S^*X$. We will denote it by $L$ in the following. Furthermore, each periodic geodesic
carries a unique invariant probability measure. The chaotic flows we will consider admit a
countable set of periodic geodesics, the union of which fills $S^*X$ densely.

Given a manifold $(X,g)$, we are led to the following question:
%
%\begin{quote}
Among all $g^t$-invariant probability measures on $S^*X$, which ones do actually appear
as semiclassical measures? Equivalently, to which invariant measures can the Wigner  measures
$(\mu_n)$ converge to in the high-frequency limit?
%\end{quote} 

At the moment, the answer to this question for a general manifold $X$ is unknown. We will
henceforth be less ambitious and restrict ourselves to geodesic flows satisfying
well-controlled dynamical properties:
%on one side, the fully integrable systems, in the sense of Liouville-Arnold-Mineur 
%\cite{Duister80}; on the opposite side, 
the \textit{strongly chaotic systems}.

\vspace{-.1\baselineskip} %\smallskip

\section{Chaotic geodesic flows}

The word ``chaotic'' is quite vague so we will need to provide more precise dynamical
assumptions. All chaotic flows we will consider are \textit{ergodic} with respect to the
Liouville measure. This means that $S^*X$ cannot be split into two invariant subsets of
positive measures. A more ``physical'' definition is the following: the trajectory starting
on a \textit{typical} point $\rho\in S^*X$ will cover $S^*X$ in a uniform way at long times
(see Figure \ref{fig:stade}) so that ``time average equals spatial average''.

The ``stadium'' billiard (see Figure \ref{fig:stade}) enjoys a stronger chaoticity:
\textit{mixing}, meaning that any (small) ball $B\subset S^*X$ evolved through the flow
will spread uniformly throughout $S^*X$ for large times. The strongest form of chaos is
reached by the geodesic flow on a manifold of negative curvature; such a flow is
\textit{uniformly hyperbolic} or, equivalently, it has the Anosov property \cite{An67}. All
trajectories are then unstable with respect to small variations of the initial conditions.
Paradoxically, this strong instability leads to a good mathematical control on the long
time properties of the flow. Such a flow is fast mixing with respect to $L$.

Numerical computations of eigenmodes are easier to perform for Euclidean billiards than on
curved manifolds; on the other hand, the semiclassical analysis is more efficient in the
case of boundary-free compact manifolds so most rigorous results below concern the latter.

\subsection{Quantum ergodicity}

Ergodicity alone already strongly constrains the structure of the high-frequency
eigenmodes: \textit{almost all} of these eigenmodes are equidistributed on $S^*X$.
\begin{thm} [\textbf{Quantum ergodicity}]\cite{Shni74,Zel87,CdV85} Assume the geodesic flow
on $(X,g)$ is ergodic with respect to the Liouville measure on $S^*X$.

Then, there exists a subsequence $(n_j)\subset \N$ of density $1$ such that the Wigner
measures of the corresponding eigenmodes satisfy
\[
\mu_{n_j}\Lim_{j\to \infty} L\,.
\]
\end{thm}
\noindent
The phrase ``of density $1$'' means that $\frac{\#\{n_j\leq
N\}}{N}\stackrel{N\to\infty}{\to}1$. Therefore, if there exists a subsequence of eigenmodes
converging towards a semiclassical measure $\mu\neq L$, this subsequence must be sparse and
consist of \textbf{exceptional eigenmodes}.

\subsection{``Scars'' and exceptional semiclassical measures}

%\begin{sloppy}

\looseness=1
Numerical computations of eigenmodes of some chaotic billiards have revealed interesting
structures. In 1984, Heller \cite{Hel84} observed that some eigenmodes of the ``stadium''
billiard (the ergodicity of which had been demonstrated by Bunimovich) are ``enhanced''
along some periodic geodesics. He called such an enhancement a ``scar'' of the periodic
geodesic upon the eigenmode (see Figure \ref{fig:stade}). Although it is well-understood
that an eigenmode
%(or at least a quasimode\footnote{A quasimode of
%precision $\eps$ is a solution of $\norm{(-\hbar^2\lap-I)\psi}\leq \eps\norm{\psi}$. The existence of
%such a quasimode implies that of a true eigenvalue in the interval  $[1-\eps, 1+\eps]$. Still, the 
%quasimode $\psi$ does not necessarily resemble any true eigenmode. In the limit $\hbar\to 0$, this
%notion is relevant only if $\eps=\eps(\hbar)$ decreases fast enough.})
can be concentrated along a \textit{stable} periodic geodesic, the observed localization
along \textit{unstable} geodesics is more difficult to justify. The enhancement observed by
 Heller was mostly ``visual''; the more quantitative studies that followed
Heller's paper (e.g.\ \cite{Barnett06}) seem to exclude the possibility of a positive
probability weight remaining in arbitrary small neighbourhoods of the corresponding
geodesic. Such a positive weight would have indicated that the corresponding semiclassical
measures ``charge'' the unstable orbit (a phenomenon referred to as ``strong scar'' in
\cite{RudSar94}). Contrary to the case of Euclidean billiards, numerical studies on
surfaces of constant negative curvature have not shown the presence of ``scars''
\cite{AurichSteiner93}.

%\end{sloppy}

On the mathematical level, the most precise results on the localization of eigenmodes are
obtained in the case of certain surfaces of constant negative curvature enjoying specific
\textit{arithmetic symmetries}, called ``congruence surfaces''. A famous example is the
modular surface (which is not compact). For these surfaces, there exists a commutative
algebra of selfadjoint operators on $L^2(X)$ (called Hecke operators), which also commute
with the Laplacian. It is then reasonable to focus on the orthonormal bases formed of joint
eigenmodes of these operators (called Hecke eigenmodes). Rudnick and Sarnak have shown
\cite{RudSar94} that semiclassical measures associated with such bases cannot charge any
periodic geodesic (``no strong scar'' on congruence surfaces). This result, as well as the
numerical studies mentioned above, suggested to them the following
\begin{conj}[\textbf{Quantum unique ergodicity}]\label{conf:QUE}
Let $(X,g)$ be a compact Riemannian manifold of negative curvature. For any orthonormal
eigenbasis of the Laplacian, the sequence of Wigner measures $(\mu_n)_{n\geq 0}$ admits a
unique limit (in the weak topology), namely the Liouville measure.
\end{conj}

\noindent
This conjecture goes far beyond the non-existence of ``strong scars''. It also excludes all
the ``fractal'' invariant measures.

This conjecture has been proved by E.~Lindenstrauss in the case of compact congruence
surfaces, provided one only considers \textit{Hecke eigenbases} \cite{Linden06}.
%and as well for the non-compact ones when combining his work with results of Soundararajan \cite{Sound}. 
The first part of the proof \cite{BLi03} (which relies heavily on the Hecke algebra)
consists of estimating from below the \textit{entropies} of the ergodic components of a
semiclassical measure. We will see below that the entropy is also at the heart of our
results.

\subsection{The role of multiplicity?\label{s:mult}}

\textit{A priori}, there can be multiple eigenvalues in the spectrum of the Laplacian, in
which case one can make various choices of orthonormal eigenbases. On a negatively curved
surface, it is known \cite{Ber77} that the eigenvalue $k_n^2$ has multiplicity
$\cO\left(\frac{k_n}{\log k_n}\right)$ but this is far from what people expect, namely a
uniformly bounded multiplicity. One could modify Conjecture~\ref{conf:QUE} so that the
statement holds for a given basis (e.g.\ a Hecke eigenbasis in the case of a congruence
surface) but may be false for another basis.

In parallel with the study of chaotic geodesic flows, people have also considered toy
models of discrete time symplectic transformations on some compact phase spaces. The most
famous example of such transformations is better known as ``Arnold's cat map'' on the
2-dimensional torus. It consists of a linear transformation $(x,\xi)\to M(x,\xi)$, where
the unimodular matrix $M\in SL(2,\mathbb{Z})$ is hyperbolic, i.e.\ it satisfies $|\tr
M|>2$. The ``Anosov property'' then results from the fact that no eigenvalue of $M$ has
modulus $1$. That transformation can be \textit{quantized} to produce a family of unitary
propagators, depending on a mock Planck parameter $\hbar_N=(2\pi N)^{-1}$, where $N$ is an
integer \cite{HB80}. Such propagators have been named ``quantum maps'' and have served as a
``laboratory'' for the study of quantum chaotic systems, both on the numerical and
analytical sides.

Concerning the classification of semiclassical measures, the ``quantized cat map'' has
exhibited unexpectedly rich features. On the one hand, this system enjoys arithmetic
symmetries, allowing one to define ``Hecke eigenbases'' and prove the quantum unique
ergodicity for such eigenbases \cite{KurRud00}. On the other hand, for some (scarce) values
of $N$ the spectrum of the quantum propagator contains large degeneracies. This fact has
been exploited in \cite{FNdB03} to construct sequences of eigenfunctions violating quantum
unique ergodicity:\ the corresponding Wigner measures $\mu_N$ converge to the semiclassical
measure
\begin{equation}\label{e:demi-balafre}
\mu=\frac12\delta_O+\frac12 L,
\end{equation}
where $L=dx\,d\xi$ is now the symplectic volume measure and $\delta_0$ is the $M$-invariant
probability measure supported on a periodic orbit of $M$.

This result shows that the quantum unique ergodicity conjecture can be \textit{wrong} when
extended to chaotic systems more general than geodesic flows. More precisely, for the ``cat
map'' the conjecture holds true for a certain eigenbasis but is wrong for another one.

Another result concerning the ``cat map'' is the following: the weight $1/2$ carried by the
scar in \eqref{e:demi-balafre} is \textit{maximal} \cite{FN04}. In particular, no
semiclassical measure can be supported on a countable union of periodic orbits. In the next
section, dealing with our more recent results on Anosov geodesic flows, we will see this
factor $1/2$ reappear.

\section{Entropic bounds on semiclassical measures}

In this section, we consider the case of a compact manifold (without boundary) of negative
sectional curvature. As mentioned earlier, the corresponding geodesic flow has many
invariant probability measures. The \textbf{Kolmogorov-Sinai entropy} associated
with an invariant measure is a number $h_{KS}(\mu)\geq 0$, defined below. We stress a few
important properties:
\begin{itemize}
\item A measure supported by a periodic trajectory has zero entropy.  
\item The maximal entropy $h_{\max}$ is reached for a unique invariant measure, 
called the Bowen-Margulis measure, of support $S^* X$.
\item  According to the Ruelle-Pesin inequality,
\be\label{e:Ruelle-Pesin}
h_{KS}(\mu)\leq \int \sum_{k=1}^{d-1}\lambda_k^+ d\mu,
\ee 
where the functions
$\lambda_1^+(\rho)\geq\lambda^+_2(\rho)\geq\cdots\geq\lambda_{d-1}(\rho)>0$, defined
$\mu$-almost everywhere, are the positive Lyapunov exponents of the flow. Equality in
\eqref{e:Ruelle-Pesin} is reached only if $\mu$ is the Liouville measure \cite{LY85}.
\item On a manifold of constant curvature $-1$, the inequality reads  $h_{KS}(\mu)\leq d-1$. 
The Bowen-Margulis measure is then equal to the Liouville measure.
\item The functional $h_{KS}$ is affine on the convex set of invariant probability measures.
% \item $h_{KS}(\mu, (g^{\alpha t}))=|\alpha| h_{KS}(\mu, (g^{ t}))$.
\end{itemize}
These properties show that the entropy provides a quantitative indication of the
\emph{localization} of an invariant measure. For instance, a positive lower bound on the
entropy of a measure implies that it cannot be supported by a countable union of periodic
geodesics. This is precisely the content of our first result.
%This is precisely the contents of our first result about the entropy of semiclassical measures:
\begin{thm}
(1) \textup{\cite{An}}  
Let $X$ be a compact Riemannian manifold such that the geodesic flow has the Anosov property. Then
every semiclassical measure $\mu$ on $S^*X$ satisfies
\[
h_{KS}(\mu)>0.
\]
(2) \textup{\cite{AN07}}\label{thm:AKN}
Under the same assumptions, let $\lambda_j^+(\rho)$ be the positive Lyapunov exponents and
$\lambda_{\max}=\lim_{t\to\infty}\frac{1}t\log \sup_{\rho\in S^*X} ||dg^t_\rho||$ be the
maximal expansion rate of the geodesic flow. Then the entropy of $\mu$ satisfies
\begin{equation}\label{e:courbure-variable}
h_{KS}(\mu)\geq  \int \sum_{k=1}^{d-1}\lambda_k^+ d\mu-\frac{d-1}2 \lambda_{\max}\,.
\end{equation}
In constant curvature $-1$, this bound reads $h_{KS}(\mu)\geq \frac{d-1}2$.
\end{thm}

\begin{coro}\cite{An} 
Let $X$ be a compact manifold of dimension $d$ and constant sectional curvature $-1$. 
Then, for any semiclassical measure $\mu$, the support of $\mu$ has Hausdorff dimension $\geq d$.
\end{coro}

In constant negative curvature, the lower bound $h_{KS}(\mu)\geq \frac{d-1}2$ implies that
at most $1/2$ of the mass of $\mu$ can consist of a scar on a periodic orbit. This is in
perfect agreement with the similar result proved for ``Arnold's cat map'' (see
\S\ref{s:mult}).

The right-hand side of \eqref{e:courbure-variable} can be negative if the curvature varies
a lot, which unfortunately makes the result trivial. A more natural lower bound to hope for
would be
\begin{equation}\label{e:optimal}
h_{KS}(\mu)\geq \frac 12 \int \sum_{k=1}^{d-1}\lambda_k^+ d\mu.
\end{equation}
This lower bound has been obtained recently by G.~Rivi\`ere for surfaces ($d=2$) of \textbf{nonpositive} curvature
\cite{Riviere08}. 
B.~Gutkin has proved an analogous result for certain quantum maps with a variable Lyapunov
exponent \cite{Gutkin08}; he also constructed eigenstates for which the lower bound is
attained.

From the Ruelle--Pesin inequality \eqref{e:Ruelle-Pesin}, we notice that  proving
the quantum unique ergodicity conjecture in the case of Anosov geodesic flows would amount
to getting rid of the factor $1/2$ in \eqref{e:optimal}.

We finally provide a definition of the entropy and a short comparison between the entropy
bound of Bourgain-Linden\-strauss \cite{BLi03} and ours.

\begin{definition} 
The shortest definition of the entropy results from a theorem due to Brin and Katok \cite{BK83}. 
For any time $T>0$, introduce a distance on $S^*X$,
\[ 
d_T(\rho, \rho')=\max_{t\in[-T/2, T/2]}d(g^t\rho, g^t\rho'),
\]
where $d$ is the distance built from the Riemannian metric. For $\eps>0$, denote by
$B_T(\rho, \eps)$ the ball of centre $\rho$ and radius $\eps$ for the distance $d_T$. When
$\eps$ is fixed and $T$ goes to infinity, it looks like a thinner and thinner tubular
neighbourhood of the geodesic segment $[g^{-\eps}\rho, g^{+\eps}\rho]$ (this tubular
neighbourhood is of radius $e^{-T/2}$ if the curvature of $X$ is constant and equal to
$-1$).

Let $\mu$ be a $g^t$--invariant probability measure on $S^*X$. Then, for $\mu$-almost every
$\rho$, the limit
\begin{equation*}
\lim_{\eps\To 0}\,\liminf_{T\To +\infty}-\frac1T\log\mu \big(B_T(\rho, \eps)\big) \\
=\lim_{\eps\To 0}\,\limsup_{T\To +\infty}-\frac1T\log\mu \big(B_T(\rho, \eps)\big)
\defeq h_{KS}(\mu, \rho)
\end{equation*}
exists and it is called the local entropy of the measure $\mu$ at the point $\rho$ (it is
independent of $\rho$ if $\mu$ is ergodic). The Kolmogo\-rov-Sinai entropy is the average
of the local entropies:
\[
h_{KS}(\mu)=\int h_{KS}(\mu, \rho)d\mu(\rho).
\]
%As $\eps$ approaches $0$, the number $h_{KS}(\mu, \eps)$ approaches a limit that we define to be $h_{KS}(\mu)$. 
%On a manifold of negative curvature, $h_{KS}(\mu, \eps)$ is independent of $\eps$
%for $\eps$ small enough.

\end{definition}

\begin{rem} 
In the case of congruence surfaces, Bourgain and Lindenstrauss \cite{BLi03} proved the
following bound on the microlocal lifts of Hecke eigenbases: for any $\rho$, and all
$\eps>0$ small enough,
\[
\mu_n\big(B_T(\rho, \eps)\big)\leq C e^{-T/9},
\]
where the constant $C$ does not depend on $\rho$ or $n$. This immediately yields that any
semiclassical measure associated with these eigenmodes satisfies $\mu(B_T(\rho, \eps))\leq
C e^{-T/9}$, which implies that \textit{any ergodic component of $\mu$ has entropy $\geq
\frac19$}.

In \cite{AN07}, we work with a different, but equivalent, definition of the entropy. On a
manifold of dimension $d$ and constant curvature $-1$, the bound we prove can be (at least
intuitively) interpreted as
\begin{equation}\label{e:hyper}
\mu_n\big(B_T(\rho, \eps)\big)\leq C\,k_n^{\frac{d-1}2}\, e^{-\frac{(d-1)T}2},
\end{equation}
where $k_n^2$ is the eigenvalue of the Laplacian associated with $\psi_n$. This bound only
becomes non-trivial for times $T>\log k_n$. For this reason, we cannot directly deduce
bounds on the \linebreak weights $\mu(B_T(\rho, \eps))$; the link between \eqref{e:hyper}
and the entropic bounds of Theorem \ref{thm:AKN} is less direct and uses some specific
features of quantum mechanics.
\end{rem}

%\printnotes

\end{document}